**Integrating Fréchet distance and AI reveals the evolutionary trajectory and origin of SARS-CoV-2**

Anyou Wang ORCID 0000-0002-4981-3606

Feinstone Center for Genomic Research, University of Memphis, Memphis, TN 38152, USA

Contact:

anyou.wang@alumni.ucr.edu

Project website

https://combai.org/ai/covidgenome/

Highlights

- A novel alignment-free system integrating Fréchet distance (Fr) and artificial recurrent neural network (RNN)
- Quantitatively reveal the evolutionary trajectory and origin of SARS-CoV-2 from big data
    - Overall evolutionary trajectory: SARS-CoV-2 shortens its genome and mutates multiple biomarkers to enhance its infectious potential, contributing to the emergence of the COVID-19 pandemic.
    - Origin: Mink

**Abstract:** A genome, composed of a precisely ordered sequence of four nucleotides (ATCG), encompasses a multitude of specific genome features like AAA motif. Mutations occurring within a genome disrupt the sequential order and composition of these features, thereby influencing the evolutionary trajectories and yielding variants. The evolutionary relatedness between a variant and its ancestor can be estimated by assessing evolutionary distances across a spectrum of genome features. This study develops a novel, alignment-free algorithm that considers both the sequential order and composition of genome features, enabling computation of the Fréchet distance (Fr) across multiple genome features to quantify the evolutionary status of a variant. Integrating this algorithm with an artificial recurrent neural network (RNN) reveals the quantitative evolutionary trajectory and origin of SARS-CoV-2, a puzzle unsolved by alignment-based phylogenetics. The RNN generates the evolutionary trajectory from Fr data at two levels: genome sequence mutations and organism variants. At the genome sequence level, SARS-CoV-2 evolutionarily shortens its genome to enhance its infectious capacity. Mutating signature features, such as TTA and GCT, increases its infectious potential and drives its evolution. At the organism level, variants mutating a single biomarker possess low infectious potential. However, mutating multiple markers dramatically increases their infectious capacity, propelling the COVID-19 pandemic. SARS-CoV-2 likely originates from mink coronavirus variants, with its origin trajectory traced as follows: mink, cat, tiger, mouse, hamster, dog, lion, gorilla, leopard, bat, and pangolin. Together, mutating multiple signature features and biomarkers delineates the evolutionary trajectory of mink-origin SARS-CoV-2, leading to the COVID-19 pandemic. Full text and detailed on https://combai.org/ai/covidgenome/

## Introduction

Since the emergence of SARS-CoV-2 in 2019, its origin and evolutionary trajectory have attracted intense attention[1–14]. Understanding the origin and evolutionary trajectory of SARS-CoV-2 and the drivers of its evolutionary trajectory will help combat the COVID-19 pandemic and prevent future pandemics.

The phylogenetics tree has been employed to study virus evolution and it has provided wealth of knowledge[6,8,15,16]. However, this conventional approach cannot generate a clear evolutionary tree from millions of SARS-CoV-2 genome sequences[17]. Importantly, the traditional method depends on alignment[8,17], which underestimates sequence variations. For example, aligning two pure A sequences of P and Q, in which P and Q hold 8 As (AAAAAAAA) and 10 As (AAAAAAAAAA), respectively, results in 100% similarity. This alignment automatically ignores two AA in Q, which causes variation in codon capacity because Q codes three AAA codons, but P only has two AAA codons. Conventional phylogenetics actually finds a limited number of minor variations inside the aligned sequences and disregards the major variations hidden in the unaligned sections, which may carry primary biological functions. Therefore, alignment-based methods fail to create an unbiased evolutionary picture of genome sequence evolution.

The alignment-based method has also been applied to identify variants marked by mutation biomarkers, and has provided useful information for the static state of qualitative mutants[4,17–22]. However, SARS-CoV-2 has accumulated more than 50,000 mutations[23] and a variant (e.g., alpha) usually carries multiple mutations. Discriminating variants based on one or two signature mutations is biased, although considerable effort has been made to select variant signatures[24,25]. More importantly, virus variants continue to mutate during dynamic evolution in both the vertical and horizontal directions. Vertically, a variant can undergo deep mutations in a certain region, such as loss of a whole marker fragment. A variant can also mutate horizontally and carry many mutation markers. For example, a variant can possess alpha, gamma, delta, and kappa markers simultaneously. It is challenging for qualitative approaches to illustrate this type of evolutionary trajectory of the SARS-CoV-2 variant and to generate the general driver of this virus evolution.

Because conventional approaches fail to provide a clear picture of the variant evolution trajectory, they are unable to generate the correct origin trajectory of this virus, leading to the controversial debate of its origin, in which bats have been mostly thought to be the origin of SARS-CoV-2[10].

Alignment-free methods have been developed[26], but they have focused on either nucleotide positions or content, and they do not have enough features to discriminate between a variant genome and a reference. More importantly, these approaches lack robustness to buffer noise from sequencing experiments, and thus, they have few applications. Take together these leave the SARS-CoV-2 evolutionary trajectory as a black box.

To elucidate the unbiased evolutionary trajectory of SARS-CoV-2, this study developed a novel alignment-free system that estimates the evolutionary divergence of a variant from its ancestor by computing its Fréchet distance (Fr)[27] of multiple unbiased genome features. Fr was implemented to compute a two-dimensional distance that accounts for both the ordered position of a genome feature and its composition. After computing the Frs of all variants, long short-term memory[28] (LSTM, a recurrent neural network architecture) models were constructed to minimize noise and build a sensitive and robust Fr-LSTM system that quantitatively revealed the evolutionary trajectory of SARS-CoV-2 at both the genome sequence and organism variant levels. At genome sequence evolution, the Fr-LSTM system revealed evolutionary trajectories of multiple genomic features. At the variant evolution, quantitative variants were defined and the evolutionary trajectory of these quantitative variants was generated. The viral infectious capacity of each feature and variant was quantitatively estimated. In addition, this Fr-LSTM system revealed the origin trajectory of SARS-CoV-2 and identified the mink coronavirus as the origin of SARS-CoV-2.

## Materials and methods

### General computational environment

All data download, processing, computations, and graphing were conducted under Linux with python 3.8 and R 3.6. Deep learning neural networks were performed using TensorFlow 2.4.0 and Scikit-learn 0.24.0.

### Data resources

The SARS-CoV-2 data were downloaded from GISAID (https://www.gisaid.org/) on July 4, 2021. A total of 2,212,864 genome sequences were downloaded. These samples were subjected to a series of filtering processes, including length (>29k), N content ≤ 10, and only ATCGN. The filtered sequences were then split into human and animal groups. The human group contained 1,128,954 genomic sequences.

## Feature selection

A variant genome contains a spectrum of genome features that differ from those of its ancestor in both order and composition. Assessing the evolutionary disparity of these genome features provides an estimation of the divergence of the variant. These features can be derived from a complete permutation of the four nucleotides (ATCG) with varying numbers of bases, such as one ($4^1$), two ($4^2$), three ($4^3$), four ($4^4$), five ($4^5$), and so on. A larger number generates a specific motif, such as 20 ($4^{20}$), for a unique primer, whereas a smaller number generates a more general feature. This study aimed to characterize the overall evolutionary status of a variant and utilized general features with one to three bases, totaling 84 genomic features. These are 4 ($4^1$) single nucleotides, 16 ($4^2$) all possible dinucleotides and 64 ($4^3$) trinucleotides as following: ‘G’,’ C', 'A', 'T', 'CG', 'GC', ‘AT’, 'TA', 'CA', 'AC', 'CT', 'TC', 'GA', 'AG', 'GT', 'TG', 'TT', 'AA', 'GG', 'CC', 'TTT', 'AAA', 'TTA', ‘TGT’, 'TTG', 'ACA', 'ATT', 'AAT', 'CTT', 'ATG', 'TAA', 'CAA', 'GTT', 'ACT', 'TGA', 'TAT', 'AAC', 'TAC', 'AGA', 'AAG', 'CTA', 'TGG', 'GTG', 'TCA', 'TGC', 'TCT', 'GAA', 'GCT', 'TTC', 'AGT', 'CTG', 'CAT', 'ATA', 'GTA', 'CAC', 'GGT', 'GAT', 'CAG', 'TAG', 'ACC', 'GCA', 'CCA', 'CCT', 'GAC', 'ATC', 'AGG', 'AGC', 'GAG', 'CTC', 'GGA', 'GTC', 'GGC', 'TCC', 'GCC', 'CGT', 'ACG', 'GGG', 'CCC', 'TCG', 'CGC', 'CGA', 'GCG', 'CGG', 'CCG'.

## Discrete Frechet distance (Fr)

The Fréchet distance measures the similarity between curves by considering the ordered locations along the curves. This study considered a feature (for example, AAA) distribution along the genome as a curve and computed the coupling Fr[27] of the feature for a variant genome (for example, EPI_ISL_601443, alpha variant) against the reference genome (NC_045512), as detailed below.

P and Q are given as feature trajectories for the reference and variant, respectively.

P={p[1], p[2],...p[n]}

Q={q[1], q[2],...q[m]}

A coupling L between P and Q is a sequence

$(P[a_1], Q[b_1]), (P[a_2], Q[b_2]), \cdot\cdot\cdot , (P[a_l], Q[b_l])$

where $a_1 = 1$, $b_1 = 1$, $a_l = n$, $b_l = m$. For all $i = 1, \cdot\cdot\cdot ,l$, $a_{i+1} = a_i$ or $a_{i+1} = a_{i+1}$, and $b_{i+1} = b_i$ or $b_{i+1} = b_{i+1}$.

Fr between Q and P is defined as

Fr(Q,P) = min{ max distance($Q[b_i]$, $P[a_i]$) for all possible couplings between P and Q}

The distance of a coupling, q ($Q[b_i]$) and p ($P[a_i]$), was computed by the two-dimensional Euclidean distance of the ordered position of a feature and its content factor, as defined below.

$$d(q,p) = \sqrt{(q1-p1)^2 + (q2-p2)^2}$$

q1 and p1: the ordered position of a feature

q2 and p2: the content factor of a feature

content factor: total positions/total length, as detailed in main text and figure 1

The Euclidean distance is always positive. This study assigns a negative sign to Fr to mark it as a deletion (loss) when the number of positions for a given feature in a variant is less than that of the reference. Automatically, a feature with a positive distance was regarded as an insertion (gain).

**Fr matrix**

Each biological variant was measured by using 84 feature Frs. These 84 Frs for the 1,128,954 filtered variants were constructed into a Fr matrix containing 84 columns and 1,128,954 rows. Each row also contained the corresponding epidemiological variables deposited in the database, such as the time tag and qualitative traits. The Fr matrix was used for all trajectory analyses in this study.

**Machine learning environment**

Many machine learning models were built in this study, but all of them were based on long short-term memory (LSTM) implemented in the Keras Sequential library under Tensorflow, which was used to build models for all machine learning throughout the entire study. All data were preprocessed using MinMaxScaler from Sklearn.

**A typical LSTM model**

A typical bidirectional LSTM model containing four layers was established in this study. The first and third layers contained 20 units, and the second layer contained 40 units. To avoid overfitting, we set dropout (0.2) for two layers for all the models in this study. The programming code was deposited on the project website[29] and github[30].

Hyperparameters were optimized by the loss function via sampling, and the weekly trajectory period was based on Fourier analysis. Loss was measured by the mean absolute error (MAE), mean squared error (MSE), mean squared error (RMSE), and $R^2$ ($R^2$_score) implemented by the Keras library. Adam was used as a model optimizer. The batch size and epochs were set to 64 and 50, respectively, for machine learning.

Two independent datasets were prepared, including test (5%) and training (95%), following the order of the time series as weekly data from 2019 to 2021. The forecasts were set for 30 days after July 4, 2021.

This typical model was modified to fit different purposes across the study, as detailed in the main text.

**84 feature ranking**

The 84 features were ranked on the basis of the final score = 2 × tau_rank + fr_rank, in which tau_rank and fr_rank are the order of absolute tau and absolute Fr median, respectively, for each feature. The tau was derived from the Kendall test as described in the main text, and the Fr median was computed by LSTM-predicted Fr.

**Quantitative variant identification**

The Identifying quantitative variants involves two key steps. UMAP[31] was used for pre-clustering samples to get pre-variants, and the members of the pre-variants were corrected using an LSTM model, as detailed in the main text. This LSTM model was modified from the typical model described above with the following modifications. MAE was used for the loss. The epoch and batch sizes were set as 150 and 20, respectively. The validation split was set to 0.1.

### Final graphing

Several final summaries were drawn using ggplot2 in R. Otherwise, they were completed using the Python software.

## Results

### Method development

To understand SARS-CoV-2 evolution, we first collected SARS-CoV-2 genome data and then developed a novel approach to reveal its evolutionary trajectory. SARS-CoV-2 genome data were downloaded from GISAID (https://www.gisaid.org/) on July 4, 2021, including all available viral genome sequences (2,212,864 samples, Materials and Methods). These samples were subjected to quality filtering, and a total of 1,128,954 filtered samples were used for downstream analyses to investigate two levels of evolutionary trajectories: genome trajectory at the sequence level and variant trajectory at the organism level (Figure 1A, Materials and Methods).

The novel approach developed in this study is an alignment-free algorithm that estimates the evolutionary divergence of a variant from its ancestor by computing the Frechet distance[27] as a measurement of the similarity between a variant and the reference. This approach involves four key steps. 1) The whole virus genome was decomposed into multiple genome features like 84 features in this study, including 4 single nucleotides, 16 dinucleotides, and 64 trinucleotides (Materials and Methods). 2) A given genomic feature (for example, AAA) is located in a series of ordered positions in the entire genome in both the variant and the reference (Figure 1B). For example, for a given reference with an 8 As sequence and a variant with a 10 As sequence, the ordered positions of the AAA feature were [0, 3] and [0, 3, 6], respectively, for the reference and variant (Figure 1B). In addition to the ordered positions, this approach also considers the feature content and measures it as a content factor (total positions/total length). For example, AAA had two positions in the 8 As reference sequence, so its content factor was 2/8 in the reference sequence, but its content factor in the 10 As variant was 3/10 because of its three positions and 10 As (Figure 1B). 3) The ordered positions of a feature are treated as curves for the reference and variant. This similarity between the two curves (reference and variant) can be measured using the Frechet distance (Fr), which computes the two-dimensional Euclidean distance of the ordered position of a feature and its content factor (Materials and Methods). For example, the Fr of AAA was 3.000417 for this variant with a 10 As sequence (Figure 1B, Materials and Methods). Following this Fr practice, this study used the genome sequence NC_045512.2[32] as the wild-type reference and computed Fr for each feature (for example, AAA) for a given variant genome. A plus and minus Fr represent feature gain and loss, respectively (Materials and Methods). 4) All 84 feature Frs of a variant form an Fr array that quantitatively describes the evolutionary similarity between this variant and the reference. Therefore, this novel approach employs multiple Frs to define the evolutionary state of a variant in a high-dimensional space.

Combining arrays for all 1,128,954 samples creates a Fr matrix containing 1,128,954 rows and 84 columns. Besides 84 Frs, the Fr matrix still carries epidemiological variables such as time tag and qualitative variant name (e.g., alpha) deposited in the database.

To examine the sensitivity of Frs in measuring mutation variations, we compared the 84 feature Frs of three groups of early merging SARS-CoV-2, including China, alpha, and global (Figure 1C). The Chinese group was collected in 2019 and these samples were close to wild-type. The alpha group was a single variant (EPI_ISL_601443) marked by alpha mutation, while the global group included all worldwide samples that were collected before July 2020. The alpha Fr was computed using a single alpha sequence, but the Fr of China and the global group were the Fr medians of the Chinese and global samples, respectively. As expected, 74 out of 84 features of the Chinese group had Fr close to 0, but 10 feature Frs were below 0 (right part in Figure 1C), indicating that they were very similar to the wild-type reference, but mutations already occurred in the 2019 Chinese samples. This suggested that Fr is a sensitive metric for detecting minor mutations. In contrast, most feature Frs for the alpha and global variants moved far away from 0, indicating that these features were dramatically mutated. In addition, most feature Frs were distributed below 0, indicating that genome deletion dominated SARS-CoV-2 evolution, even in the early emerging stage.

Together, these results suggest that the novel system measuring multiple Frs (84 here) is a sensitive metric to quantitatively define the evolutionary state of a virus.

**Global evolutionary trajectory of SARS-CoV-2 genome**

To understand the global evolutionary trajectory of the SARS-CoV-2 genome, we examined the daily dynamic trend of the Fr median for all 84 features across all worldwide samples (Figure 2). These samples have been generated under various conditions, and the sequence noise and the impact of confounding variables are unavoidable. To diminish these effects and appreciate a clear trend of viral evolutionary trajectory from December 2019 to July 4, 2021, and forecast its trends beyond July 4, 2021, this study employed long short-term memory (LSTM) to find a clear trajectory (Materials and Methods). The starting point for training this model was set to 21 days (input width =21), and the forecast days were set to 30 days after July 4, 2021. The walk-forward strategy of one position shift was applied during modeling. The median Fr was used to train and predict the Fr value, and the predicted Fr, instead of the raw Fr value, was used as the metric to explain all results in this study without specific notice.

The Fr median gradually declined below 0 during the entire evolutionary trajectory (Figure 2A), indicating the general and gradual loss of its genome as a key hallmark of SARS-CoV-2 evolution. The overall genome showed three major losses during the entire trajectory. The first came from March 2020 to May 2020, with Fr sharply declining from -50 to -170. The second occurred from the middle of July 2020 to September 2020, with Fr dropping from -170 to -225. The third was the longest one that happened from November 2020 to March 2021, with Fr ranging from -225 to -280. SARS-CoV-2 will continue to shorten its genome in the near future, as forecasted after July 2021. Therefore, SARS-CoV-2 has evolved to shorten its genome length.

To estimate the infectious capacity of SARS-CoV-2, this study used the same LSTM model described above, but a multivariate (84 features) matrix was used as the training matrix, and the number of infection cases was used as the response. The predicted cases fit the actual case very well (Figure 2B), and the dynamic infection cases corresponded to the evolutionary trajectory of the virus genome (Figure 2B vs. Figure 2A). Interestingly, the predicted cases after May 2021 were much higher than actual cases, suggesting that vaccination may knock down the natural cases, and genome sequences after 5/2021 may be biased to unvaccinated patients.

To understand the relationship between genome mutations and infection cases at the global level, this study examined the correlation between the global Fr median and global infection cases using LOESS regression and Kendall test (Figure 2C). This correlation was not linear. Slight genome loss (Fr median > -75) did not change the viral infectious capacity, whereas excessive mutations (Fr median < -210) reduced the viral infectious ability. However, the wide range of genome loss, with Fr median ranging from -75 to -210, significantly increased SARS-CoV-2 infectious capacity. The overall correlation between infection cases and genome loss was also significant (tau= -0.6454225 and p-value= 1.39e-101, Kendall test, Figure 2C), indicating that SARS-CoV-2 gradually shortens its genome to enhance its infectious capacity.

The global evolutionary trajectory of SARS-CoV-2 displays a gradual depletion pattern, which drastically enhances its infectious capacity, leading to the COVID-19 pandemic.

**Evolutionary trajectories of 84 individual features**

To appreciate the dynamic evolution of the 84 genome features, this study used LSTM to model the evolutionary trajectory of each individual feature, similar to the global evolutionary trajectory above. The two trajectories of GCT and TAA highlighted the different feature evolutions during the COVID-19 pandemic (Figure 3A-3B). GCT had three sharp drops, representing three large deletion events in its evolutionary trajectory (Figure 3A), while TAA gradually inserted its contents during the evolution of this virus (Figure 3B). All 84 feature trajectories are shown in the section of "PLots of evolutionary trajectories of 84 genome features" on the project website[29].

To appreciate the full picture of these 84 feature evolutions, we calculated the Fr median of each feature and plotted these 84 Fr medians. Fr medians of 66 features were below 0, from TCG, CGT, CCC, CGA, CCA, and CG to TCA, whereas only 18 Fr medians were above 0, such as TTA and GGG (Figure 3C). This indicates that SARS-CoV-2 has deleted most of its features during evolution. Furthermore, all single nucleotides and dinucleotides were in the deletion group, with a Fr median < 0 (Figure 3C). These results further confirm the general loss of the genome pattern of SARS-CoV-2, as observed in the global pattern above. These results suggest that deletion is the primary driver process for virus evolution and also indicates the robustness of our system.

To understand the infectious capacity of each feature, this study investigated the correlation between the daily median Fr of each feature and its infectious cases. This correlation was examined using LOESS regression and a Kendall test that created the tau value and p-value for each feature. Two correlations between GCT and TTA exemplified the effect of a feature mutation on viral infectious capacity (Figure 3D-3E). Large deletions of GCT, from Fr -300 to -400, dramatically increased virus infectious capacity, from 1e+05 to 6e+05 (Figure 3D), while TAA insertion gradually enhanced virus infectious capacity (Figure 3E). All feature correlations are shown in the section of “Regression plots of 84 genome features vs their infection capacity” on the project website[29].

To understand the full profiling of infectious capacities associated with feature mutations, this study plotted the tau values of all 84 features and found that 63 feature Fr were negatively correlated with infection cases (tau <0), whereas only 21 feature Fr (for example, TTA) were positively correlated to infection cases (Figure 3F). When absolute tau = 0.4 was used as the most significant cutoff, there were a total of 38 features with tau < -0.4, and only 10 features with tau >0.4 (Figure 3F). These results indicate that most feature deletions help the virus enhance its infectious capacity and feature deletion as the dominant feature to increase infection.

To determine the signature features with both significant mutation and high infectious capacity, we ranked the absolute Fr median and tau value separately and then combined their ranking score to the final score, creating a final ranking (Figure 3G, Materials and Methods). TAA, GCT, CG, CTA, and CAT were ranked as the top five signature features. These features were significantly associated with the virus infectious capacity (absolute tau > 0.6, p-value < 9.4e-89, Figure 3G and our websites[29]), in which inserting TTA and deleting GCT and CG enhanced infectious potential. This suggests that these top-ranked features work as signature features driving the virus evolution linked to the COVID-19 pandemic.

Together, SARS-CoV-2 primarily deletes most of its genomic features during evolution and mutates signature features such as TAA, GCT, and CG to significantly enhance its infectious capacity, causing the COVID-19 pandemic.

**Quantitative variant identification**

After analyzing the evolutionary trajectories in the genome sequence, our focus shifted to exploring the trajectories of the genetic variants. We first need to identify quantitative variants, hereafter referred to simply as "variants." Because variants evolve following time series, we partitioned the dataset of 1,128,954 samples into weekly intervals, spanning from December 2019 to July 4, 2021, maintaining chronological time-series order. These weekly chunks were then utilized in a two-step process for identifying quantitative variants: pre-classification using UMAP[31], followed by correction using LSTM[28], as described below.

Here, we exemplify the details of these two steps in identifying the first five variants. In the classification process, a variant was defined as having a minimum of 50 members during the UMAP pre-classification step. To ensure an adequate number of members for each variant, we aggregated the samples from the initial eight weeks and employed UMAP to classify them into distinct clusters, resulting in five clusters labeled from 0 to 4 (Figure 4A, Materials and Methods).

These five clusters were treated as preliminary variants (pre-variants), and the members of each pre-variant cluster required further correction, with outlier members identified and removed from their respective pre-variant clusters. The correction process involved an LSTM model with four layers (Materials and Methods). The LSTM model sequentially corrected the variant membership for each pre-variant from 0 to 4, as exemplified below.

To determine the members of pre-variant 0, the LSTM model used all members of pre-variant 0 as the training set, whereas members of the remaining pre-variants (pre-variants 1-4) served as the test set. The mean absolute error (MAE) was calculated from the LSTM training set members (see Figure 4B), and the threshold for filtering members was set as the MAE mean plus 1.5 times the standard deviation. Members with an MAE below the threshold were considered corrected final members for variant 0, whereas those with an MAE above the threshold were filtered out as outliers (see Figure 4C), resulting in final variant 0. The outliers were reintroduced into the sample pool for subsequent UMAP cycles.

This process was iteratively applied to correct the members of pre-variants 1 to 4, maintaining consistency with the procedure outlined above (see Figure 4B-4C). These processes finally identified 5 variants from 0 to 4.

To visualize the evolutionary state of these five variants, we plotted a heatmap of these five variants with all 84 feature Frs (Figure 4D). This heatmap showed that variant 0 was close to the reference sequence with few feature alternations (Figure 4D), but variant 2-4 had undergone a series of mutations. Variant 4 had already mutated most of its 84 features (Figure 4D) in 8 week data. This rapid evolution of variants indicated that SARS-CoV-2 rapidly mutated once it adapted to human immunity and that our system was sensitive enough to discriminate variants. This variant diversity emerging within eight weeks also suggests that SARS-CoV-2 might remain in the human community for a long time before being reported.

After the first five variants were identified, the weekly window was moved to identify the next group of variants. Following the same algorithm and practice, 34 variants were identified (Figure 4E). These 34 variants and their compositions were used for the analysis of variant evolution.

**Vertical and horizontal mutations in quantitative variants**

A variant can undergo both vertical and horizontal mutations during its evolution. During vertical mutation, a variant mutates a certain region heavily (e.g., losing the alpha qualitative marker region), whereas a variant can also mutate horizontally, in which it mutates multiple sites across its entire genome to gain multiple markers, leading to high richness of markers. To obtain a snapshot of vertical and horizontal mutation profiling of each quantitative variant, this study examined its member composition by decomposing its members into 12 qualitatively classified groups, such as alpha, which was officially defined by the World Health Organization(WHO) by July 4, 2021[21](Figure 4E). The profiling of vertical and horizontal mutations in all 34 variants is summarized in Figure 4E and below.

Variant 0 only contained 100% of the unknown category and it had a richness of 1 (out of 12), indicating that variant 0 was close to the wild type and underwent few vertical and horizontal mutations. In contrast, variant 23 comprised 82% of the alpha group, indicating that it vertically mutated its genome, which was heavily characterized by an alpha group mutation. On the other hand, a total of four variants (variants 13, 17, 21, and 24) carried all 12 qualitative markers defined by the WHO, indicating that they underwent heavy horizontal mutation and possessed the highest richness of 12 (Figure 4E). These vertical and horizontal mutations and richness empower variants with various biological significances, as discussed below.

**Evolutionary trajectories of variants**

To understand the evolutionary trajectories of these 34 variants, this study employed the LSTM model and replaced only the input for training and testing with variant data. The evolutionary trajectories of these 34 variants showed diversity, as detailed in the section of "SARS-CoV-2 variant evolutionary trajectories" on this project website[29]. For example, variant 13 only underwent an intermediate level of mutation with Fr median > -300, and it underwent three waves of deletion, beginning in 06/2020, 1/2021, and 04/2021(Figure 5A). These deletion waves corresponded to the three waves of global viral infection outbreaks. In addition, variant 13 possessed 12 qualitative markers in the horizontal mode, as described above, suggesting that variant 13 gradually mutates to extend its mutation marker richness. In contrast, variant 23 had the largest deletion among all variants with Fr near -600 (Figure 5B), and it displayed the sharpest and largest deletion in 1/2021 (Fr = -650). Its deletion recovered slightly after 03/2021, but its Fr was always low (<-550) (Figure 5B). In addition, variant 23 underwent a vertical mutation, as shown above. This indicated that variant 23 had the largest vertical deletion of alpha markers during the entire evolution.

To appreciate the overlook of these 34 variants, we plotted the Fr median for all 34 variants and revealed that 32 out of 34 variants suffered the overall deletion with Fr median < 0 (Figure 5C) and only two variants (variant 0 and 1) almost behaved like wild type (Fr = ~0). These results suggest that horizontal and vertical deletions predominate SARS-CoV-2 evolution, and further confirm that SARS-CoV-2 dynamically deletes its genome during its evolution in the human community. These horizontal and vertical mutations affect various infectious capacities, as described below.

**Infectious trajectories of SARS-CoV-2 variants**

With the wavering evolutionary trajectories of SARS-CoV-2 variants, as shown above, a variant can dynamically change its infectious capacity along the evolutionary trajectory. To appreciate the infectious capacity trajectory of each variant, this study first trained an LSTM model with a global viral 84 feature matrix versus known global infection cases as done above (Figure 2B) and kept the parameters of this trained LSTM model, and then used the 84 feature Fr matrix of a variant (e.g., variant 13) to fit this trained LSTM model to estimate the infectious capacity for the variant (e.g., variant 13).

The whole infectious trajectories of 34 variants are shown in the section of “SARS-CoV-2 variant infection capacity” on our website[29]. Here, we used variants 13 and 23 as examples to demonstrate the infection trajectory (Figure 5D-5E). Variants 13 and 23 were predicted to infect a maximum 550k and 140k cases per day, respectively (Figure 5D-5E). As they evolved in their genome, variant 13 increased its infectious capacity along its evolutionary trajectory, whereas variant 23 maintained its low infectious capacity all time.

To compare the infectious capacity across all variants associated with their vertical mutation levels, we used the alpha proportion listed in the alpha column in Figure 4E to represent the vertical mutation level for each variant. We plotted the variant LOESS regression between the alpha proportion and maximum daily infection cases and found a correlation pattern between vertical mutations and their capacities (Figure 5F). First, variants with a few vertical mutations showed little infectious potential. For example, variant 0 caused few infections (Figure 5F). However, increasing vertical mutation levels from 0 to intermediate gradually enhanced virus infectious capacity, but over-mutations at the vertical level dramatically reduced the virus infectious potential (Figure 5F). For example, variant 13, with an intermediate mutation, had a higher infectious potential than variant 23, which carried the highest vertical mutation (Figure 5F).

To understand the association between horizontal viral mutations and infectious potential, we plotted the LOESS regression between maximum daily infection cases and the richness of qualitative markers (Figure 5G, Figure 4E). Variant richness was positively correlated with infectious potential (Figure 5G). In particular, variants with the highest richness (variants 17, 21, 13, and 24) possessed the highest infectious capacity, indicating that mutating diverse markers is an important metric for viral infectious potential. Together, these results suggest that horizontal mutations, instead of vertical mutations, are the most important drivers of the COVID-19 pandemic.

**SARS-CoV-2 origin path**

To understand the origin of SARS-CoV-2, this study examined the distance from the animal coronavirus to SARS-CoV-2. The distance was measured using MAE derived from LSTM, as described above (Figure 4B). Before running LSTM, human SARS-CoV-2 samples were reversed in a time series from 2021 to 2019 to trace the origin of SARS-CoV-2. Human variant 0 identified above (Figure 4) served as the wild-type (WT). The Fr matrix of WT was used as a training set to fit the LSTM model, and animal samples were treated as a test to calculate MAE for samples as described above (Figure 4B). Animal samples with the lowest MAE were close to the human wild type. Ranking the minimum MAE for all animals revealed that mink was very close to human wild type (MAE near 0, Figure 6A), followed in order by cat, tiger, mouse, hamster, dog, lion, gorilla, leopard, bat, and pangolin. This indicates that the mink coronavirus has the ability to infect humans directly and is the most likely origin of SARS-CoV-2. In contrast, it is unlikely that coronaviruses from bats or pangolins directly infect humans.

To understand how mink coronavirus was so close to SARS-CoV-2, we plotted the MAE between human WT and mink (Figure 6B) and found that several of the mink samples mutated to be similar to human WT. Moreover, these mink mutants had 56 consensus features (defined as those with the same sign of positive and negative Fr, Figure 6C), and they had 25 features (out of 56) different from those of normal mink viruses (Figure 6D). These mink viruses actually shared 57% (32 out of 56) features with humans (Figure 6E), while only 16% (9/56) were different between minks and humans. Therefore, the mink mutants were predicted to be of SARS-CoV-2 origin.

## Discussion

In this study, we developed a novel alignment-free algorithm designed to capture unbiased evolutionary relatedness among the genomes of organisms. Traditionally, phylogenetic similarity based on sequence alignment has been utilized to estimate such evolutionary relationships. However, the conventional approach typically emphasizes minor alignment-filtered variations, resulting in biased representations. The algorithm developed in this study aims to estimate the evolutionary relatedness between a variant and its ancestor by evaluating the similarity of all provided genome features. The similarity of a genome feature is assessed using Fr, which considers both feature order and composition. Furthermore, all genome features are derived without bias from the complete permutation of the four nucleotides (ATGC), with varying numbers of bases, such as one ($4^1$), two ($4^2$), three ($4^3$), four ($4^4$), five ($4^5$), six ($4^6$), and so forth. Consequently, this alignment-free algorithm produces an unbiased depiction of the evolution of an organism.

The present study computed 84 feature Frs to characterize a SARS-CoV-2 variant, and these 84 Frs were sensitive enough to discriminate a variant with only 30k length. However, a future application can certainly expand it to additional features, such as ($4^4 + 4^5 + 4^6 + 4^7$), to enhance the discrimination specificity and sensitivity when genomes become complex. A recent study used motifs $4^6$ and $4^7$ to select biologically significant motifs in the human genome[33].

Assembling all variant Frs constructs an Fr matrix. This Fr matrix can be computed using a range of algorithms such as AI transformer models, LSTM models, conventional machine learning models, and even simple statistical methods. For example, it can serve as an input for generating an evolutionary tree through traditional clustering algorithms. In this study, LSTM models were utilized to extract clear evolutionary trajectories from extensive datasets characterized by high levels of noise stemming from diverse sequence sources and confounding variables. These LSTM models effectively reduce noise and pinpoint essential trajectories, thereby strengthening the robustness of the Fr system for this specific dataset, which spans daily data points over approximately one and a half years. Nevertheless, it is apparent that besides LSTM, other algorithms can also detect similar patterns within this Fr matrix.

Consequently, this Fr measurement can be computed using various algorithms for evolutionary studies, particularly those involving large datasets.

The evolutionary trajectory of the SARS-CoV-2 genome has been widely regarded as one of the most crucial topics in SARS-CoV-2 studies[5,6,8,34], yet it remains undisclosed. This study systematically revealed a quantitative evolutionary trajectory of the SARS-CoV-2 genome. Generally, SARS-CoV-2 gradually shortens its genome to enhance its capacity for infection, although the relationship between genome loss and viral infectious capacity is not linear. Among the 84 features analyzed, it primarily deleted 66 features, such as CG and GCT, while it only gained 18 features, such as TTA. These feature mutations significantly increase the infectious capacity of the virus and act as drivers propelling its evolution. A recent study also reported CG deficiency in the viral genome[35], but our results revealed an insightful trajectory. In wild-type SARS-CoV-2 (before March 2021), CG deficiency is not severe, but it is only lost during evolution (after March 2020), as detailed on our website[29].

The evolutionary trajectory of SARS-CoV-2 variants has been intensively reviewed and discussed[5,6]; however, this has not yet been resolved. This study employed LSTM and Fr to identify 34 quantitative SARS-CoV-2 variants and revealed their evolutionary trajectories. SARS-CoV-2 has undergone both vertical and horizontal mutations during evolution. Vertical mutation helps a variant increase its infectious capacity, but over-mutation dramatically reduces its infectious potential. Vertically mutated variants possess low infectious capacity and are unlikely to cause pandemics. In contrast, horizontal mutations increase mutation marker diversity, which helps a variant dramatically increase its infectious capacity and works as a driver of the COVID-19 pandemic.

The origin of SARS-CoV-2 has been the subject of debate, and its path remains a mystery even after years of study[6,8,11–13,36–38]. Although certain animals such as bats and pangolins have been suspected to be potential viral sources[39,40], the complete path involving multiple intermediate hosts has yet to be definitively established. This study sheds light on the origin trajectory of SARS-CoV-2, which unfolds as follows: mink, cat, tiger, mouse, hamster, dog, lion, gorilla, leopard, bat, and pangolin. Recent studies have demonstrated viral transmission between animals and human[40,41], such as instances where cat coronaviruses can infect humans[42] and mink coronaviruses can transinfect humans[7]. Therefore, SARS-CoV-2 likely originated from animals in close proximity, such as minks, cats, and mice, rather than from bats and pangolins, as previously suggested.

This study has several limitations. First, although the algorithm developed in this study permits the utilization of a variable number of genome features derived from permutations of the four nucleotides (ATGC), it is important to acknowledge that handling a large number of features requires significant computational resources and time. Furthermore, a large number, such as 20, typically used for generating PCR primers unique to specific genes, would produce a plethora of unique primers that could not be found in many variants. Consequently, it is crucial to plot the density distribution of feature Frs across various numbers to determine an appropriate threshold based on the genome complexity. Second, while the study illuminated the evolutionary trajectory at the genome level, it did not delve into specific genes and proteins. Future endeavors should aim to apply this algorithm when ample data on specific gene functions become available, enabling exploration of the functional evolution of given genes and proteins.

This study not only uncovers how SARS-CoV-2 evolves to infect humans, but also provides a novel and reliable system to study the evolution of any organism. This system can be employed to more accurately predict future pandemics.

## Declarations

### Ethics approval and consent to participate

NA

### Consent for publication

NA

### Availability of data and material

All detailed figures and data was deposited in the project website[29]

### Competing interests

NO

### Funding

No funding resource associated with this project

### Authors' contributions

AW for All

**Acknowledgment**

Thank GISAID (https://www.gisaid.org) for providing the full data.

## Figure legends

**Figure 1. Computing the Fr of the genome features.** A, workflow. This study investigated SARS-CoV-2 evolutionary trajectories at both the genome sequence level (upper path) and the organism variant level (lower path). At the genome sequence level, the evolutionary trajectories of 84 genome features were revealed, whereas at the variant level, this study identified quantitative variants and investigated the evolutionary trajectory of each quantitative variant. B, the principle and algorithm used to compute the Fr. Three Frs of typical features (A, AA, and AAA) are exemplified here for a variant with a sequence of 10 As, and a reference with a sequence of 8 As. The second and third columns show the ordered positions and content factors of the features. For example, AAA has an ordered position [0, 3] and a content factor of 2/8. The fourth column shows Fr and its computational formula. C, Total 84 feature Fr comparisons between three virus groups: Alpha, China, and global. Alpha Fr denotes the Fr of a single virus (EPI_ISL_601443/B117), but China and global Fr represent the Fr median of Chinese samples collected in 2019 and all global samples collected before July 2020 respectively. This plot followed the order of the Chinese medians.

**Figure 2. Global evolutionary trajectory of the SARS-CoV-2 genome and its infectious capacity**. A, The global evolutionary trajectory of the SARS-CoV-2 genome, which was presented by the Fr median of LSTM prediction across worldwide samples, with MAE, MSE, RMSE, and R2 of 0.01535, 0.00039, 0.01998, and -1.35965, respectively. The Fr color scheme for actual, prediction, test, test prediction, and forecast is applied to all figures below. B, global infection cases and LSTM-predicted cases. The outliers on the right end indicate the vaccine effect. C, Loess regression (green line) between Fr median and infection cases, with tau = 0.645 and p-value =1.3e-101 (Kendall test). The Fr median and infection cases were the LSTM-predicted values from A and B, respectively. All statistical values were LSTM-predicted unless specified otherwise.

**Figure 3. 84 genome feature trajectories and infectious capacities.** Evolutionary trajectories of the two features: GCT (A) and TTA (B). All Feature trajectories are shown on the project website[29]. C, distribution of the predicted Fr median of the 84 genome features. Green, red, and blue colors denote the features of a single nucleotide, dinucleotide, and trinucleotide, respectively. D-E, Loess regression (green line) between Fr median and infection cases for GCT (D) and TTA (E). F, distribution of 84 feature tau derived from the Kendall test. The horizontal red line denotes the cut-off value of absolute tau = 0.4. G, 84 feature rankings based on both absolute tau and absolute Fr medians (Materials and Methods).

**Figure 4. Quantitative variant identification.** A, UMAP classifies the first 8 week samples into five clusters, including 0, 1, 2, 3, and 4 clusters. These five clusters were treated as five pre-variants labeled 0 to 4. B, MAE distribution for pre-variant 0. The MAE mean + 1.5 standard deviation was set as the threshold to detect outliers. C, Outlier detection for the variant 0. Members above the red line

(threshold) were treated as outliers and removed from variant 0 and members below the red line were corrected as the final members of variant 0. Thus, correcting pre-variant 0 was identified as variant 0. E, heat map of five variants with 84 feature frames. F, members and composition of 34 variants. All members of each variant were decomposed into 12 qualitatively classified groups such as alpha, gamma, delta, and unknown. For example, variant 0 contained 871 members that were 100% in the unknown group, whereas variant 23 contained 3347 members with 82% alpha samples and many other groups.

**Figure 5. Evolutionary trajectories of variants.** A-B, evolutionary trajectories of variants 13 (A) and 23 (B). C, Fr median of the 34 variants. D-E, infectious capacity trajectories of variants 13 (D) and 23 (E). The evolutionary trajectories of all variants and their infection capacities are shown on the project website[29]. F, LOESS regression (span=0.1) between maximum infection cases and the alpha proportion of the 34 quantitative variants. These 34 variants were ranked by their alpha composition proportions that were listed in the alpha column in Figure 4E. G, LOESS regression(span=0.1) between maximum infection cases and variant richness, which was used to rank the 34 quantitative variants.

**Figure 6. SARS-CoV-2 Origin.** A, origin path of SARS-CoV-2 based on the minimum MAE ranking. B, human and mink variant MAE distributions. C, top 56 consensus features of the top three mink samples close to humans. D, comparison of 56 features between the top three mink samples and the total mink samples. E, comparison of 56 features between the top three mink samples and the top five human samples.

Fig.1

A Workflow

Genome evolution → 84 genome features → Frechet distance → AI → Genome trajectory

2 M sequences → 1M sequences

Variant evolution → Weekly batches → Pre-variant clusters → Variant clusters → variant trajectory and origin

B Computating Frechet distance (Fr)

| Feature | ref:AAAAAAAA | variant: AAAAAAAAAA | Fréchet distance |
|---|---|---|---|
| A (position and content) | [0,1,2,3,4,5,6,7], 8/8 | [0,1,2,3,4,5,6,7,8,9], 10/10 | 2 = abs(9-7) |
| AA | [0,2,4,6], 4/8 | [0,2,4,6,8], 5/10 | 2 = abs(8-6) |
| AAA | [0,3], 2/8 | [0,3,6], 3/10 | 3.000417 = sqrt((6-3)^2+(3/10-2/8)^2) |

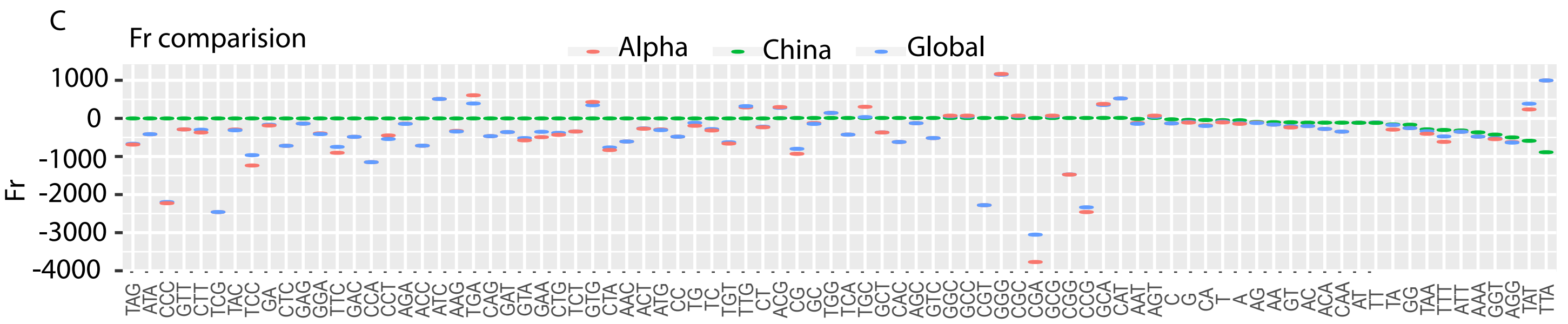

Fig.2

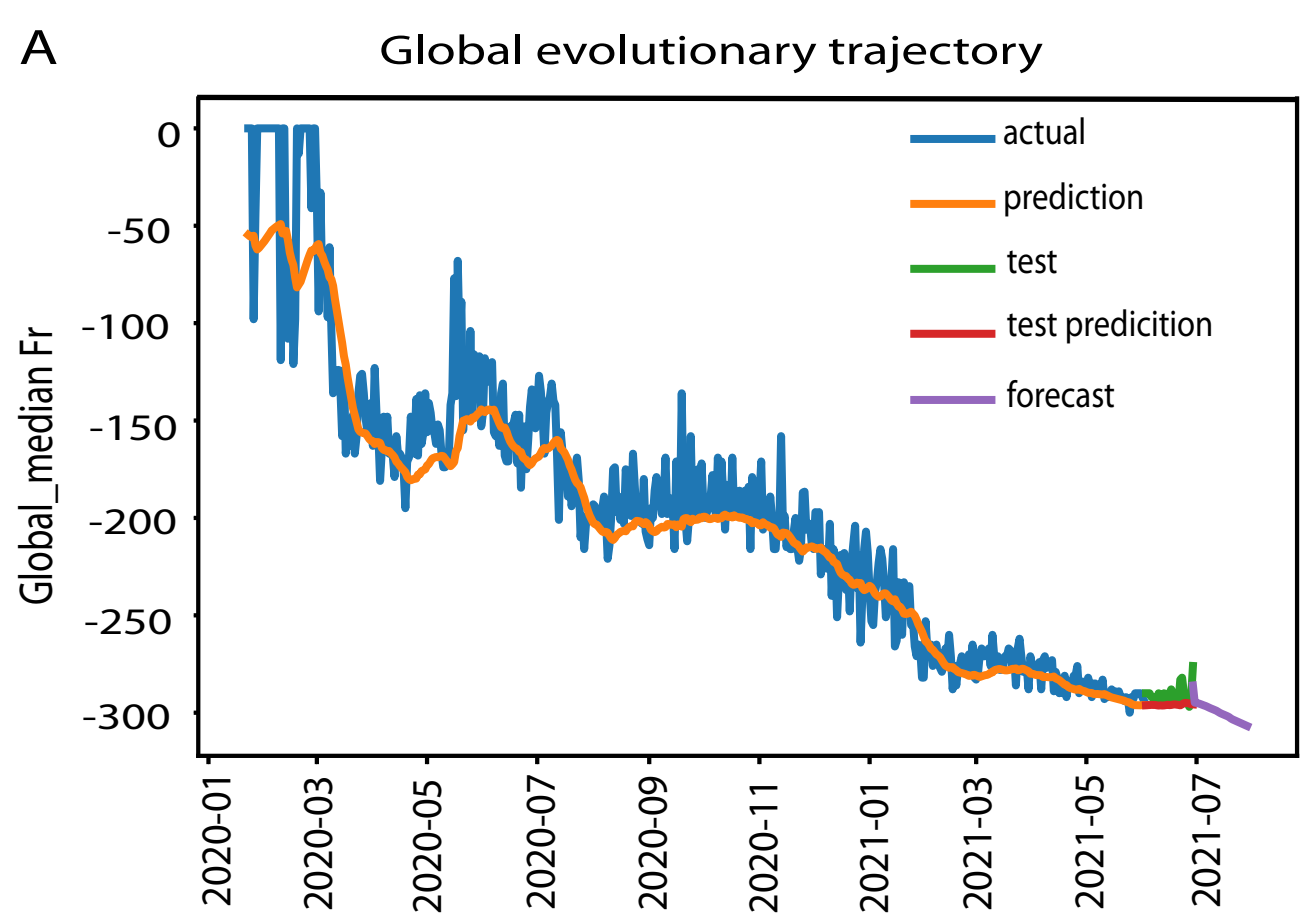

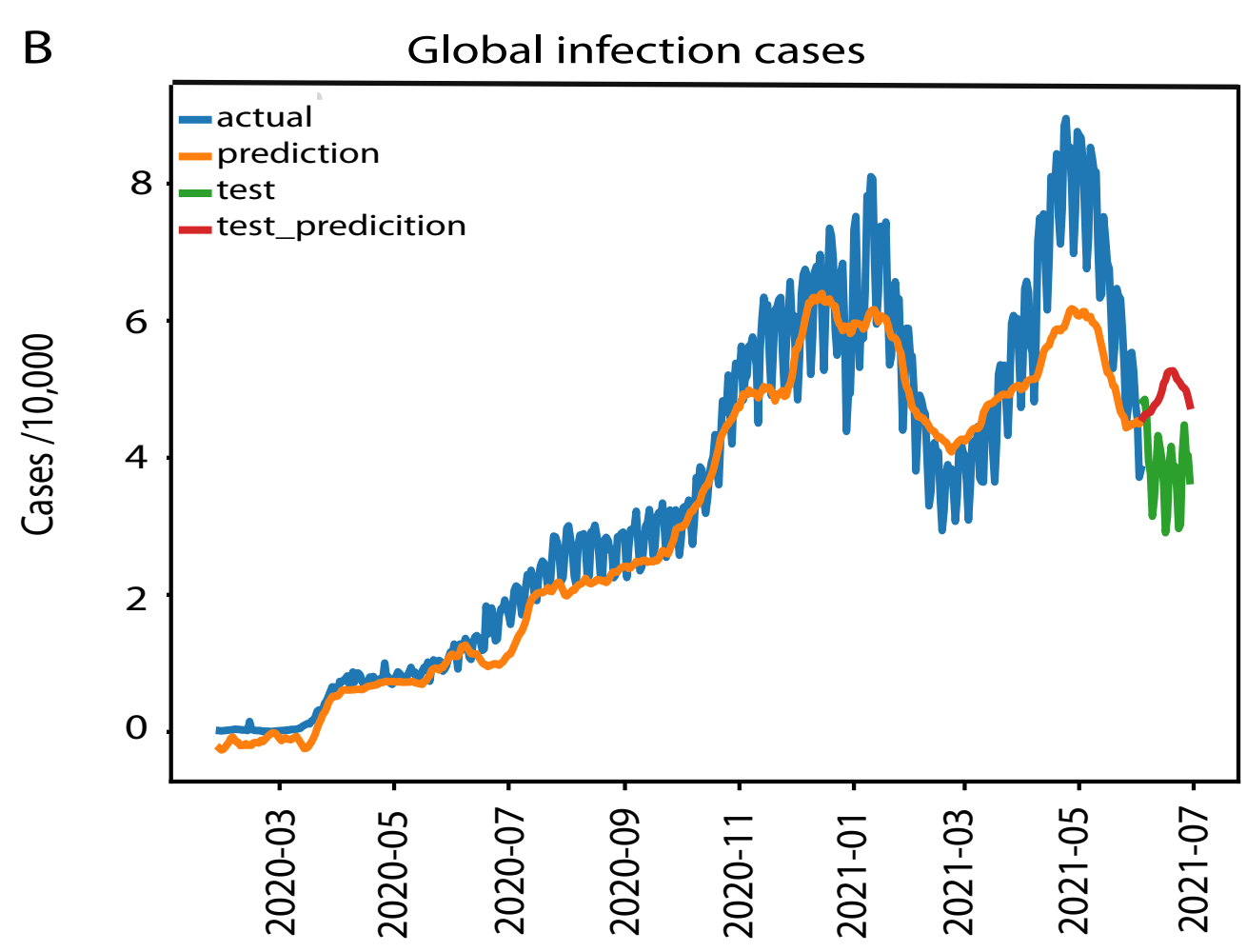

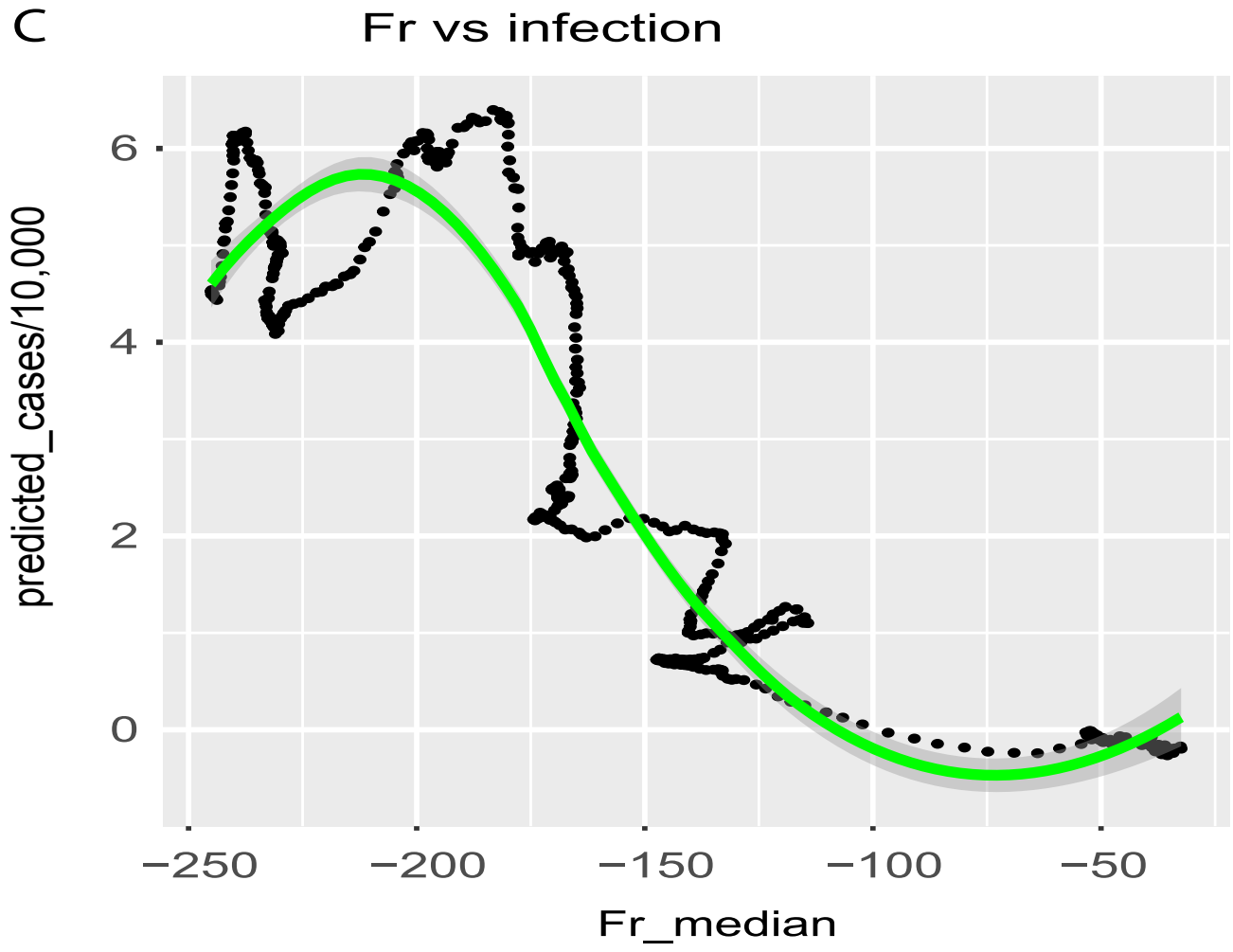

Fig.3

A GCT evolutionary trajectory

Fr median

B TTA

C 84 feature Fr median

Fr median

di
single
tri

D GCT Fr vs infection cases

Infection cases

Fr

E TTA Fr vs infection cases

Fr

F Distribution of 84 feature tau

tau

di
single
tri

G

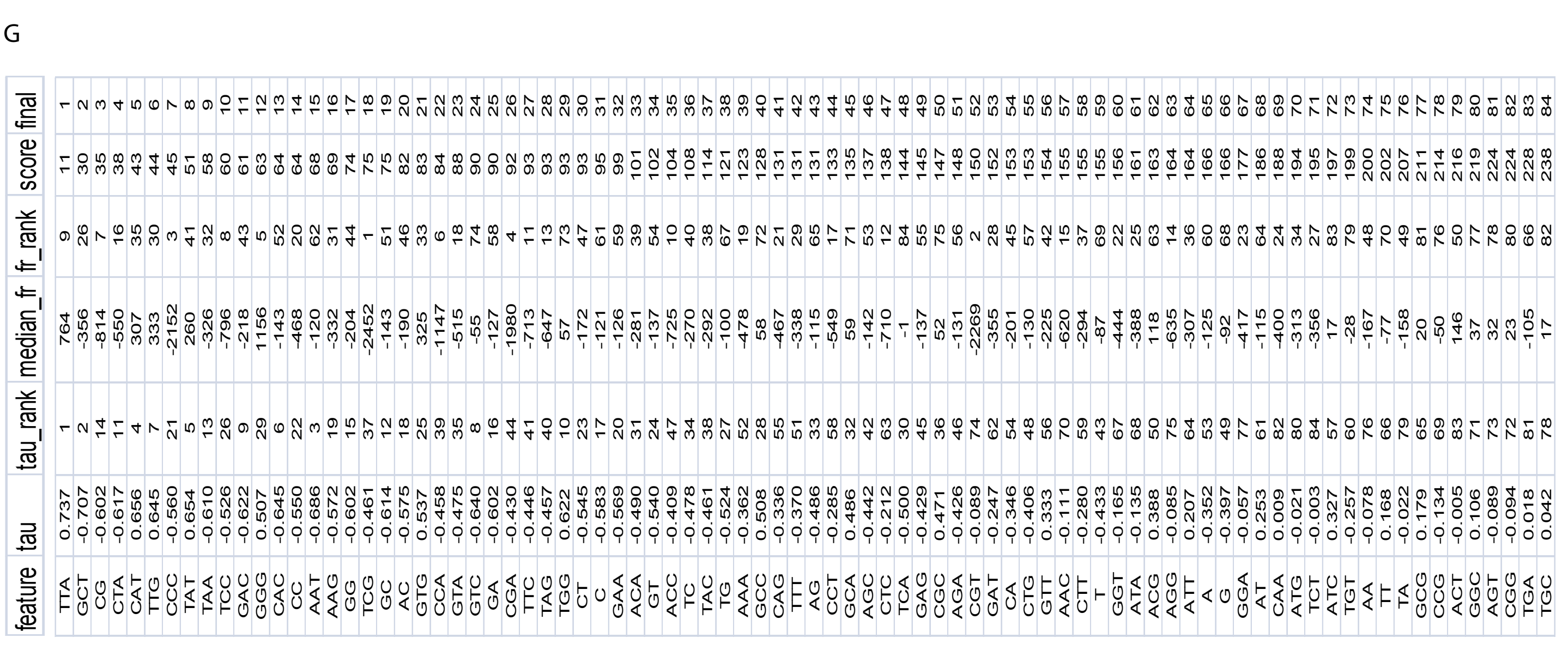

| feature | tau | tau_rank | median_fr | fr_rank | score | final |
|---|---|---|---|---|---|---|
| TTA | 0.737 | 1 | 764 | 9 | 11 | 1 |
| GCT | -0.707 | 2 | -356 | 26 | 30 | 2 |
| CG | -0.602 | 14 | -814 | 7 | 35 | 3 |
| CTA | -0.617 | 11 | -550 | 16 | 38 | 4 |
| CAT | 0.656 | 4 | 307 | 35 | 43 | 5 |
| TTG | 0.645 | 7 | 333 | 30 | 44 | 6 |
| CCC | -0.560 | 21 | -2152 | 3 | 45 | 7 |
| TAT | 0.654 | 5 | 260 | 41 | 51 | 8 |
| TAA | -0.610 | 13 | -326 | 32 | 58 | 9 |
| TCC | -0.526 | 26 | -796 | 8 | 60 | 10 |
| GAC | -0.622 | 9 | -218 | 43 | 61 | 11 |
| GGG | 0.507 | 29 | 1156 | 5 | 63 | 12 |
| CAC | -0.645 | 6 | -143 | 52 | 64 | 13 |
| CC | -0.550 | 22 | -468 | 20 | 64 | 14 |
| AAT | -0.686 | 3 | -120 | 62 | 68 | 15 |
| AAG | -0.572 | 19 | -332 | 31 | 69 | 16 |
| GG | -0.602 | 15 | -204 | 44 | 74 | 17 |
| TCG | -0.461 | 37 | -2452 | 1 | 75 | 18 |
| GC | -0.614 | 12 | -143 | 51 | 75 | 19 |
| AC | -0.575 | 18 | -190 | 46 | 82 | 20 |
| GTG | 0.537 | 25 | 325 | 33 | 83 | 21 |
| CCA | -0.458 | 39 | -1147 | 6 | 84 | 22 |
| GTA | -0.475 | 35 | -515 | 18 | 88 | 23 |
| GTC | -0.640 | 8 | -55 | 74 | 90 | 24 |
| GA | -0.602 | 16 | -127 | 58 | 90 | 25 |
| CGA | -0.430 | 44 | -1980 | 4 | 92 | 26 |
| TTC | -0.446 | 41 | -713 | 11 | 93 | 27 |
| TAG | -0.457 | 40 | -647 | 13 | 93 | 28 |
| TGG | 0.622 | 10 | 57 | 73 | 93 | 29 |
| CT | -0.545 | 23 | -172 | 47 | 93 | 30 |
| C | -0.583 | 17 | -121 | 61 | 95 | 31 |
| GAA | -0.569 | 20 | -126 | 59 | 99 | 32 |
| ACA | -0.490 | 31 | -281 | 39 | 101 | 33 |
| GT | -0.540 | 24 | -137 | 54 | 102 | 34 |
| ACC | -0.409 | 47 | -725 | 10 | 104 | 35 |
| TC | -0.478 | 34 | -270 | 40 | 108 | 36 |
| TAC | -0.461 | 38 | -292 | 38 | 114 | 37 |
| TG | -0.524 | 27 | -100 | 67 | 121 | 38 |
| AAA | -0.362 | 52 | -478 | 19 | 123 | 39 |
| GCC | 0.508 | 28 | 58 | 72 | 128 | 40 |
| CAG | -0.336 | 55 | -467 | 21 | 131 | 41 |
| TTT | -0.370 | 51 | -338 | 29 | 131 | 42 |
| AG | -0.486 | 33 | -115 | 65 | 131 | 43 |
| CCT | -0.285 | 58 | -549 | 17 | 133 | 44 |
| GCA | 0.486 | 32 | 59 | 71 | 135 | 45 |
| AGC | -0.442 | 42 | -142 | 53 | 137 | 46 |
| CTC | -0.212 | 63 | -710 | 12 | 138 | 47 |
| TCA | -0.500 | 30 | -1 | 84 | 144 | 48 |
| GAG | -0.429 | 45 | -137 | 55 | 145 | 49 |
| CGC | 0.471 | 36 | 52 | 75 | 147 | 50 |
| AGA | -0.426 | 46 | -131 | 56 | 148 | 51 |
| CGT | -0.089 | 74 | -2269 | 2 | 150 | 52 |
| GAT | -0.247 | 62 | -355 | 28 | 152 | 53 |
| CA | -0.346 | 54 | -201 | 45 | 153 | 54 |
| CTG | -0.406 | 48 | -130 | 57 | 153 | 55 |
| GTT | 0.333 | 56 | -225 | 42 | 154 | 56 |
| AAC | -0.111 | 70 | -620 | 15 | 155 | 57 |
| CTT | -0.280 | 59 | -294 | 37 | 155 | 58 |
| T | -0.433 | 43 | -87 | 69 | 155 | 59 |
| GGT | -0.165 | 67 | -444 | 22 | 156 | 60 |
| ATA | -0.135 | 68 | -388 | 25 | 161 | 61 |
| ACG | 0.388 | 50 | 118 | 63 | 163 | 62 |
| AGG | -0.085 | 75 | -635 | 14 | 164 | 63 |
| ATT | 0.207 | 64 | -307 | 36 | 164 | 64 |
| A | -0.352 | 53 | -125 | 60 | 166 | 65 |
| G | -0.397 | 49 | -92 | 68 | 166 | 66 |
| GGA | -0.057 | 77 | -417 | 23 | 177 | 67 |
| AT | 0.253 | 61 | -115 | 64 | 186 | 68 |
| CAA | 0.009 | 82 | -400 | 24 | 188 | 69 |
| ATG | -0.021 | 80 | -313 | 34 | 194 | 70 |
| TCT | -0.003 | 84 | -356 | 27 | 195 | 71 |
| ATC | 0.327 | 57 | 17 | 83 | 197 | 72 |
| TGT | -0.257 | 60 | -28 | 79 | 199 | 73 |
| AA | -0.078 | 76 | -167 | 48 | 200 | 74 |
| TT | 0.168 | 66 | -77 | 70 | 202 | 75 |
| TA | -0.022 | 79 | -158 | 49 | 207 | 76 |
| GCG | 0.179 | 65 | 20 | 81 | 211 | 77 |
| CCG | -0.134 | 69 | -50 | 76 | 214 | 78 |
| ACT | -0.005 | 83 | 146 | 50 | 216 | 79 |
| GGC | 0.106 | 71 | 37 | 77 | 219 | 80 |
| AGT | -0.089 | 73 | 32 | 78 | 224 | 81 |
| CGG | -0.094 | 72 | 23 | 80 | 224 | 82 |
| TGA | 0.018 | 81 | -105 | 66 | 228 | 83 |
| TGC | 0.042 | 78 | 17 | 82 | 238 | 84 |

Fig.4

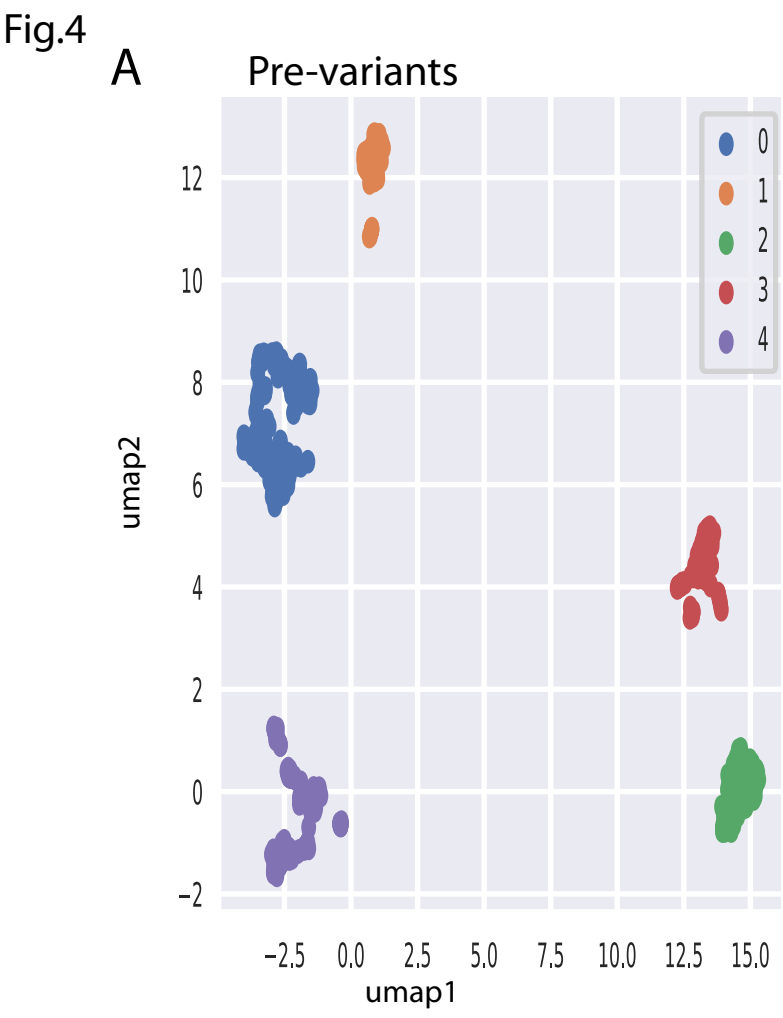

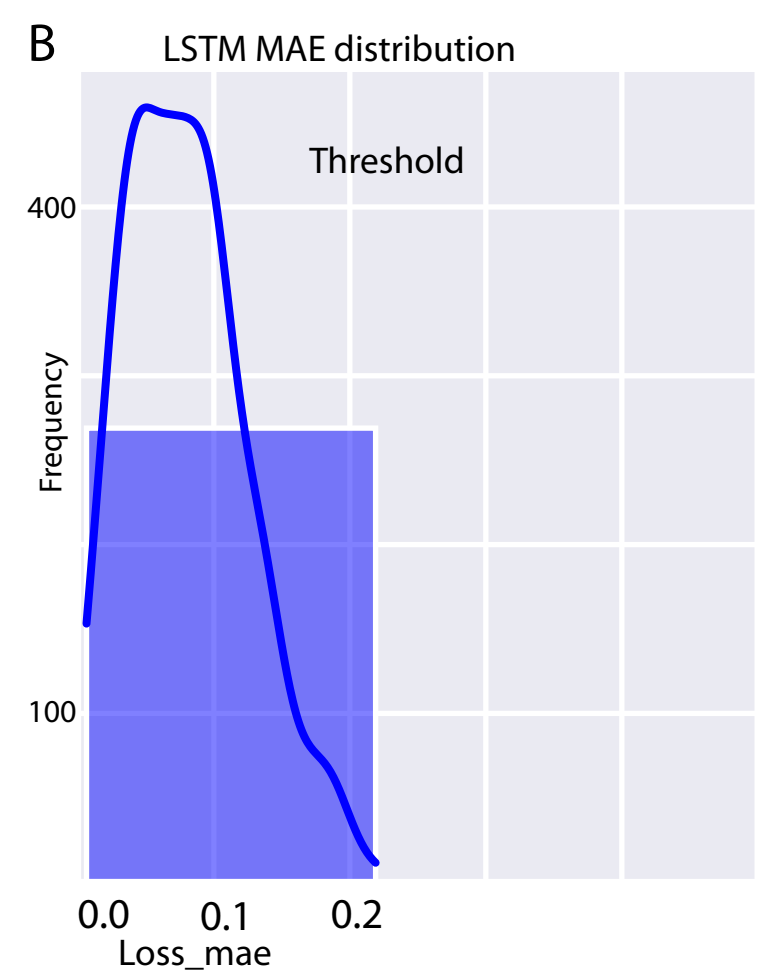

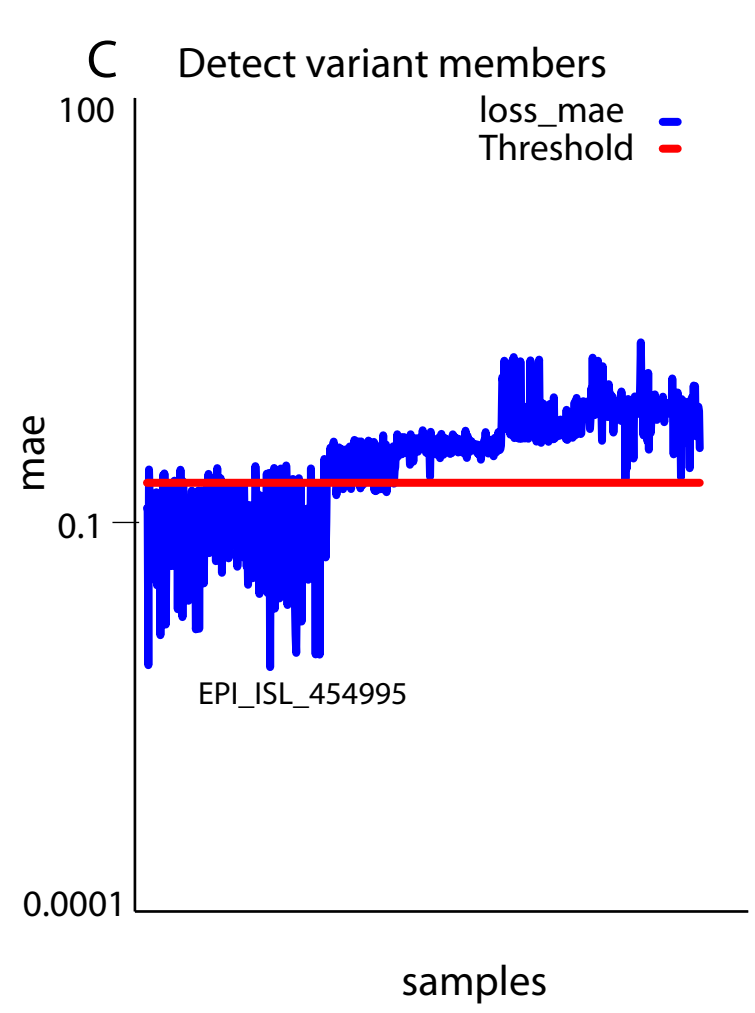

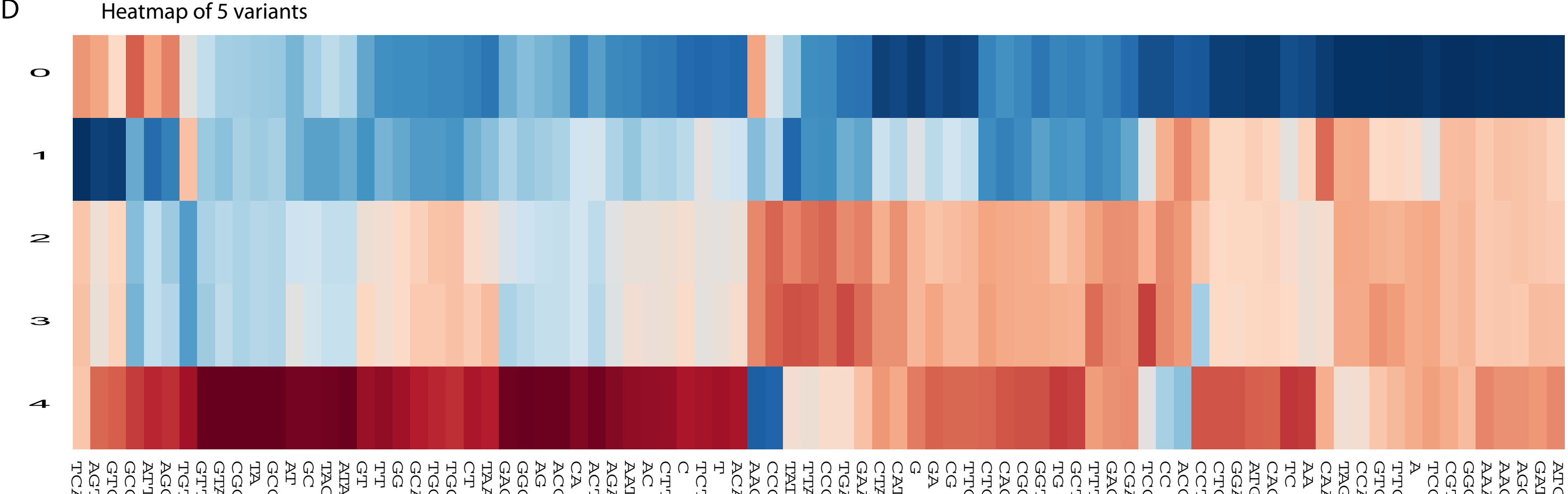

E 34 variants and their compositions

| Variant_ | member_number | Alpha | Gamma | Delta | unknown_ | Kappa | Theta | Beta | Epsilon | Zeta | Eta | Lambda | Iota |
|---|---|---|---|---|---|---|---|---|---|---|---|---|---|
| 0 | 871 | 0 | 0 | 0 | 1 | 0 | 0 | 0 | 0 | 0 | 0 | 0 | 0 |
| 1 | 272 | 0 | 0 | 0 | 1 | 0 | 0 | 0 | 0 | 0 | 0 | 0 | 0 |
| 2 | 435 | 0 | 0 | 0 | 1 | 0 | 0 | 0 | 0 | 0 | 0 | 0 | 0 |
| 3 | 1564 | 0.00064 | 0 | 0 | 0.99808 | 0 | 0 | 0.000639 | 0 | 0 | 0 | 0 | 0.00064 |
| 4 | 38916 | 0.00046 | 0.00069 | 0.000103 | 0.99733 | 7.7E-05 | 0 | 5.14E-05 | 0.000257 | 0.000694 | 5.1393E-05 | 0 | 0.00028 |
| 5 | 23623 | 0.00093 | 0 | 8.47E-05 | 0.95602 | 0.00089 | 0 | 0.000593 | 0.000212 | 0.000339 | 0 | 0 | 0.04093 |
| 6 | 822 | 0 | 0 | 0 | 1 | 0 | 0 | 0 | 0 | 0 | 0 | 0 | 0 |
| 7 | 2477 | 0 | 0 | 0 | 0.9996 | 0 | 0 | 0 | 0.000404 | 0 | 0 | 0 | 0 |
| 8 | 3920 | 0 | 0 | 0 | 1 | 0 | 0 | 0 | 0 | 0 | 0 | 0 | 0 |
| 9 | 11767 | 0.00059 | 0 | 0 | 0.99813 | 0 | 0 | 0 | 8.5E-05 | 0 | 0 | 0 | 0.00119 |
| 10 | 148492 | 0.31758 | 6.7E-06 | 6.06E-05 | 0.66179 | 3.4E-05 | 0 | 0.000431 | 0.002916 | 0.000929 | 0.00086873 | 5.387E-05 | 0.01533 |
| 11 | 7867 | 0.00127 | 0 | 0.000127 | 0.98551 | 0 | 0 | 0.01144 | 0 | 0.001271 | 0 | 0 | 0.00038 |
| 12 | 799 | 0 | 0 | 0 | 1 | 0 | 0 | 0 | 0 | 0 | 0 | 0 | 0 |
| 13 | 101535 | 0.14942 | 0.00853 | 0.010381 | 0.64263 | 0.00718 | 2.95E-05 | 0.016684 | 0.017954 | 0.012498 | 0.00521987 | 0.0009947 | 0.12849 |
| 14 | 3415 | 0.24422 | 0.01347 | 0.000293 | 0.60791 | 0 | 0 | 0.02284 | 0.000586 | 0 | 0.01083455 | 0 | 0.09985 |
| 15 | 507 | 0 | 0 | 0 | 0.97633 | 0 | 0 | 0 | 0.021696 | 0.001972 | 0 | 0 | 0 |
| 16 | 3657 | 0.01531 | 0.0175 | 0.019962 | 0.44736 | 0.00547 | 0 | 0.003555 | 0.329505 | 0.003828 | 0 | 0.0013672 | 0.15614 |
| 17 | 495215 | 0.64872 | 0.00962 | 0.051493 | 0.26108 | 0.00124 | 3.43E-05 | 0.004174 | 0.007506 | 0.001737 | 0.00225559 | 0.0004422 | 0.0117 |
| 18 | 22664 | 0.01275 | 0.00106 | 0.013766 | 0.86304 | 0.00419 | 0 | 0.033886 | 0.048888 | 0.02087 | 4.4123E-05 | 0.0006177 | 0.00088 |
| 19 | 5002 | 0.03778 | 0.0112 | 0.008597 | 0.76389 | 0.0004 | 0 | 0.003798 | 0.097161 | 0.002199 | 0.0019992 | 0 | 0.07297 |
| 20 | 216 | 0 | 0 | 0 | 1 | 0 | 0 | 0 | 0 | 0 | 0 | 0 | 0 |
| 21 | 38730 | 0.54312 | 0.0181 | 0.0189 | 0.32342 | 0.00124 | 7.75E-05 | 0.012187 | 0.052388 | 0.002505 | 0.00560289 | 0.0017816 | 0.02068 |
| 22 | 5564 | 0.44123 | 0.14216 | 0.016894 | 0.37203 | 0 | 0 | 0.00647 | 0.010963 | 0.003954 | 0.00017973 | 0.0008986 | 0.00521 |
| 23 | 3347 | 0.82133 | 0.00418 | 0.067224 | 0.09352 | 0 | 0 | 0.001494 | 0.007768 | 0.000598 | 0.00029878 | 0 | 0.00359 |
| 24 | 200524 | 0.74046 | 0.06382 | 0.057838 | 0.11266 | 0.00046 | 1.99E-05 | 0.001696 | 0.018437 | 0.000618 | 0.00079292 | 0.0007131 | 0.00248 |
| 25 | 3681 | 0.56072 | 0.13366 | 0.006248 | 0.17903 | 0 | 0.000272 | 0.001358 | 0.07335 | 0.004618 | 0.01657158 | 0.0038033 | 0.02037 |
| 26 | 396 | 0 | 0 | 0 | 1 | 0 | 0 | 0 | 0 | 0 | 0 | 0 | 0 |
| 27 | 314 | 0.00637 | 0 | 0 | 0.98408 | 0 | 0 | 0 | 0.003185 | 0.003185 | 0 | 0.0031847 | 0 |
| 28 | 113 | 0 | 0 | 0 | 1 | 0 | 0 | 0 | 0 | 0 | 0 | 0 | 0 |
| 29 | 604 | 0 | 0 | 0.001656 | 0.63576 | 0.00166 | 0 | 0.004967 | 0.35596 | 0 | 0 | 0 | 0 |
| 30 | 894 | 0.01678 | 0.01454 | 0.011186 | 0.9094 | 0.00112 | 0.001119 | 0.016779 | 0.010067 | 0 | 0 | 0.0167785 | 0.00224 |
| 31 | 251 | 0 | 0 | 0 | 0.94024 | 0 | 0 | 0 | 0.059761 | 0 | 0 | 0 | 0 |
| 32 | 379 | 0.01583 | 0.05805 | 0.055409 | 0.76517 | 0 | 0 | 0.01847 | 0.044855 | 0 | 0 | 0.0395778 | 0.00264 |
| 33 | 116 | 0.27586 | 0.02586 | 0.017241 | 0.63793 | 0 | 0 | 0.025862 | 0.017241 | 0 | 0 | 0 | 0 |

Fig.5

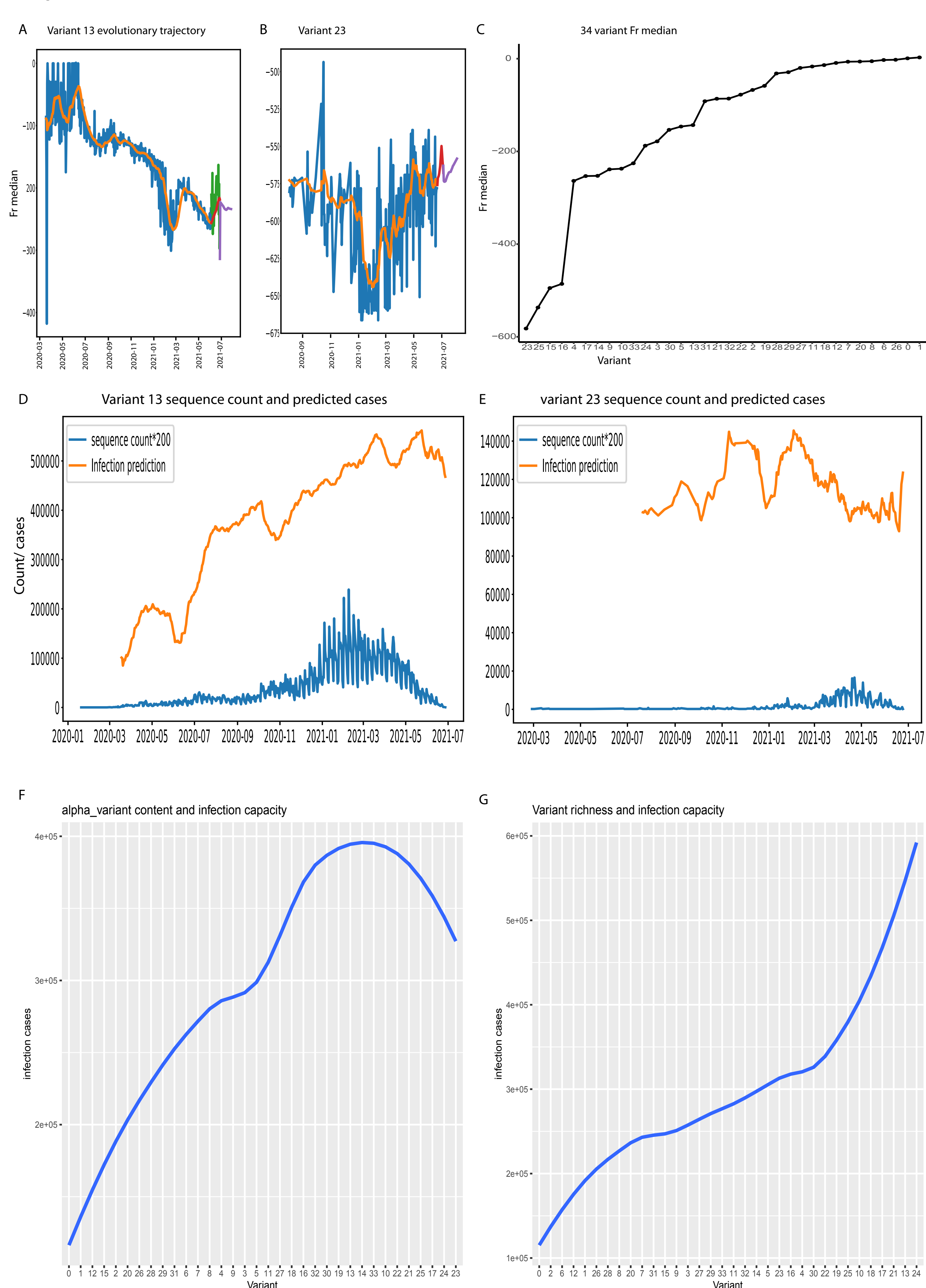

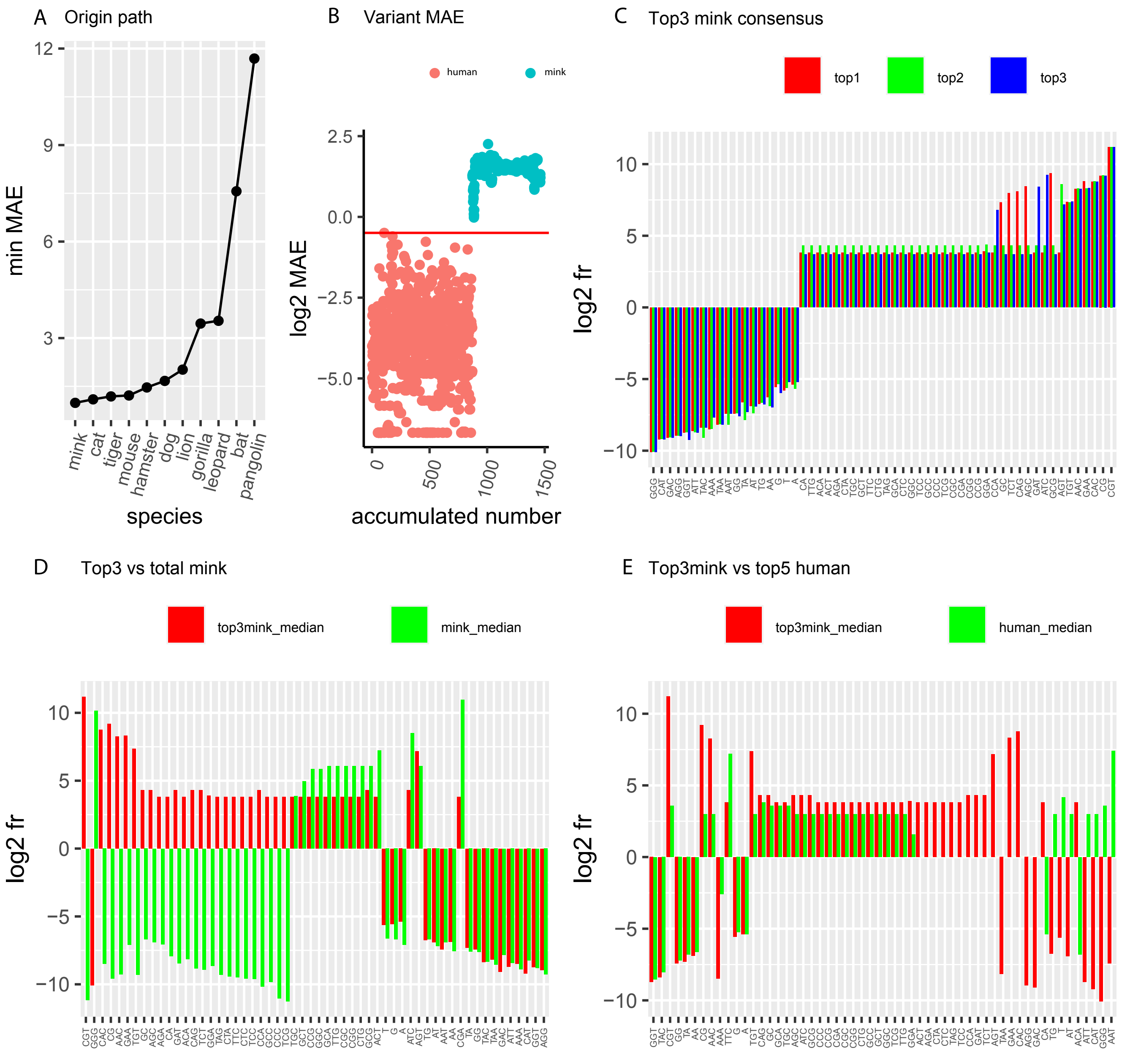